\documentclass[twocolumn]{aastex62}

\usepackage{placeins}

\begin{document}

\title{Carbon Abundance Inhomogeneities and Deep Mixing Rates in Galactic 
Globular Clusters}
\author{Jeffrey M. Gerber}
\affiliation{Department of Physics and Astronomy, Appalachian State 
University, 525 Rivers Street\\Boone, NC 28608 USA}
\affiliation{Department of Astronomy, Indiana University Bloomington, Swain West 319, 727 East 3rd Street, Bloomington, IN 47405-7105, USA}
\email{jemigerb@indiana.edu}

\author{Michael M. Briley}
\affiliation{Department of Physics and Astronomy, Appalachian State 
University, 525 Rivers Street\\Boone, NC 28608 USA}
\affiliation{Visiting Astronomer, Kitt Peak National Observatory, National 
Optical Astronomy Observatory, which is operated by the Association of 
Universities for Research in Astronomy (AURA) under cooperative agreement with 
the National Science Foundation.}
\affiliation{Visiting astronomer, Cerro Tololo Inter-American Observatory, 
National Optical Astronomy Observatory, which are operated by the Association 
of Universities for Research in Astronomy, under contract with the National 
Science Foundation. The SOAR Telescope is a joint project of: Conselho Nacional
de Pesquisas Cient,ficas e Tecnol,gicas CNPq-Brazil, The University of North 
Carolina at Chapel Hill, Michigan State University, and the National Optical 
Astronomy Observatory.}
\email{brileymm@appstate.edu}

\author{Graeme H. Smith}
\affiliation{Visiting Astronomer, Kitt Peak National Observatory, National 
Optical Astronomy Observatory, which is operated by the Association of 
Universities for Research in Astronomy (AURA) under cooperative agreement with 
the National Science Foundation.}
\affiliation{University of California Observatories/Lick Observatory, 
University of California, Santa Cruz, CA 95064 USA}
\email{graeme@ucolick.org}

\begin{abstract}
Among stars in Galactic globular clusters the carbon abundance tends to 
decrease with increasing luminosity on the upper red giant branch, particularly
within the lowest metallicity clusters. While such a phenomena is not predicted
by canonical models of stellar interiors and evolution, it is widely held to be
the result of some extra mixing operating during red giant branch ascent which 
transports material exposed to the CN(O)-cycle across the radiative zone in
the stellar interior and into the base of the convective envelope, whereupon 
it is brought rapidly to the stellar surface.

Here we present measurements of [C/Fe] abundances among 67 red giants in 19 globular 
clusters within the Milky Way. Building on the work of Martell et al., we have concentrated on giants with absolute
magnitudes of $M_\mathrm{V} \sim -1.5$ within clusters encompassing a range of metallicity (-2.4 $<$ [Fe/H] $<$ -0.3). The Kitt Peak National Observatory (KPNO) 4 m and Southern Astrophysical Research (SOAR) 4.1 m telescopes were used to obtain
spectra covering the $\lambda$4300 CH and $\lambda$3883 CN bands. The CH 
absorption features in these spectra have been analyzed 
via synthetic spectra in order to obtain [C/Fe] abundances. These abundances 
and the luminosities of the observed stars were used to infer the rate at which
C abundances change with time during upper red giant branch evolution (i.e., the 
mixing efficiency). By establishing rates over a range of metallicity, the 
dependence of deep mixing on metallicity is explored. We find that the 
inferred carbon depletion rate decreases as a function of metallicity, although
our results are dependent on the initial [C/Fe] composition assumed for each
star.
\end{abstract}


\section{Introduction}

It is known from a number of observational studies
that as low-mass (0.5 $\leq$ $M_\odot$ $\leq$ 2.0), metal-poor stars in globular clusters (GCs) 
evolve up the red giant branch (RGB) and reach
absolute magnitudes brighter than $M_\mathrm{V} \sim 0$ their surface carbon abundance ([C/Fe])
declines markedly with continued increasing luminosity \citep[see][and references therein]{gratton}.
This phenomenon (first identified in GCs, but later observed in field stars as well as open clusters; see e.g., \citealt{keller2001,szigeti2018}) defies canonical stellar models that predict surface abundances
of evolved giants remain static after the first dredge-up event.
Once metal-poor giants evolve beyond this point, a radiative zone separates the 
hydrogen-burning shell and the surface, and canonical models typically do not
allow for mass motion within this zone. 
This has lead to the suggestion of several possible extra mixing mechanisms 
that could enable the transport of CN(O) exposed material to the base of
the convective envelope, e.g., \cite{Sweigart} who developed a model of
deep meridional circulation within the 
radiative zone induced by stellar rotation. Other drivers of a deep mixing 
process such as turbulent diffusion \citep{Den},
or Rayleigh-Taylor \citep{Eggleton} or thermohaline instabilities 
\citep{Charb,Den} also have been also considered.
Theoretical modeling studies of extra mixing in metal-poor red giants
include those of \citet{denissenkov1996}, \citet{denissenkov2000},
\citet{weiss2000}, \citet{denissenkov2003},
\citet{palacios2006}, \citet{suda2006}, \citet{cantiello2010},
\citet{angelou2011,angelou2012}, \citet{lagarde2012}, and
\citet{henkel2017}.

However, common to all theories of extra mixing is a metallicity
sensitivity to its efficiency (extra mixing becoming less efficient
with metallicity) and the expectation
that extra mixing only occurs in stars that have evolved beyond a local 
maximum in the RGB luminosity function of GCs. This local maximum
(denoted herein as the LMLF) is a result of a stutter in evolution
when the 
hydrogen-burning shell advances outwards and encounters a 
significant gradient in mean 
molecular weight (a $\mu$-barrier) that formed during
the first dredge-up event.\footnote{Prior to the first 
dredge-up, the convective envelope of an evolving GC star moves inward in mass 
as the core contracts. The inward movement of this envelope results in 
partially processed material from the stellar interior mixing with 
unprocessed surface material, producing a molecular weight 
discontinuity at the base of the envelope.
As hydrogen burning continues in a shell 
around the core, the temperature gradient near the core increases and 
eventually forces the base of the convective envelope outwards. A $\mu$-barrier
is left behind at the greatest point of inward progress of the convective 
envelope \citep{Iben}.} When the hydrogen-burning shell eventually encounters 
the $\mu$-barrier, the sudden influx of hydrogen-rich material 
causes a short-term reversal in the luminosity evolution of the star. In 
a collection of GC giants having the same ages and initial compositions, the 
effect is to produce a local maximum in the luminosity function. Once a 
star has evolved beyond the RGB LMLF, any inhibiting effects of the 
$\mu$-barrier to mass transport across the radiative zone have
been removed.
The location of the LMLF is also a function of metallicity.
For higher metallicity stars, the base of the convective envelope is driven 
lower during the first dredge-up phase, which means that the hydrogen-burning 
envelope will reach the $\mu$-barrier earlier in the evolution up the RGB. 
Because the $\mu$-barrier is reached earlier in the evolution of the star, the 
LMLF will therefore occur at a lower luminosity for clusters with higher 
metallicities \citep{Zoccali}.

Here we have estimated the carbon depletion rate as a function of 
metallicity for GC stars that have surpassed the LMLF event in their evolution
up the RGB. Stars were observed from multiple Milky 
Way GCs across a range in [Fe/H] metallicity.
Using a similar method to \citet{Martell}, a carbon depletion 
rate for each star has been derived based on the measured [C/Fe] abundance,
an assumed initial abundance, and the time since each star evolved through 
the LMLF event, the latter being found through the use of isochrones for each 
cluster. The observations and results are discussed below.

\section{Observations}

The goal of this project was to study the carbon abundances of evolved red 
giants chosen to be of comparable absolute magnitude and selected from multiple
GCs encompassing a wide range of metallicities. Spectra were
obtained for 67 stars chosen on the basis of various photometric surveys 
for 19 GCs. Observations of the majority of the stars (60) were obtained
at Kitt Peak National Observatory (KPNO) with the Mayall 4 m telescope, while
11 additional stars were observed using the Southern Astrophysical Research 
Telescope (SOAR). The program stars were chosen to have an absolute 
magnitude of $M_\mathrm{V} \sim -1.5$ in order to guarantee that they were sufficiently 
evolved beyond the LMLF of the RGB of the parent cluster, but 
still distinguishable from asymptotic giant branch (AGB) stars.

\subsection{KPNO Data}

The observations made at KPNO used the 4 m Mayall telescope with the R-C 
Spectrograph under NOAO Program 2010A-0388. This system employs the 
UV Fast camera to transmit light to a 2048$\times$2048 Tektronix (T2KB) 
CCD detector with 24 $\mu$m pixels. The dispersive element used was the KPC-007
grating (632 l mm$^{-1}$ in first order blazed at 5200 \AA), which produces a 
nominal inverse dispersion of 1.4 \AA\ pixel$^{-1}$ at the CCD detector for a 
resolution of 4 \AA\ FWHM. Wavelength coverage ranged from the violet 
cutoff to $\sim$ 5500 \AA, and so includes the 
$\lambda$4300 CH band and the $\lambda$3883 CN band. The spectrograph was used
in the long-slit mode, with a width of 2'' (300 $\mu$m). Each 
program star was placed at approximately the same location on the slit. For 
each star, usually 1-2 exposures were taken for total exposure times of 
900-1800 s. The resultant spectra had a typical
signal-to-noise ratio (S/N) of $\sim$ 40 per pixel just blueward of the CH feature. Calibrations, including observations of a He, Ar, Ne comparison arc before or after each exposure, biases, and flats, followed standard practices.
The observations were made over a five-night period from 2010 June 10 to 14.
Weather was mostly clear. A sample spectrum is shown as the top panel in Figure \ref{sample}.

\subsection{SOAR Data}

\begin{figure}[htbp]
\centering
\includegraphics[trim = 1.6cm 0.4cm 0.4cm 0.4cm, scale=0.48, clip=True]{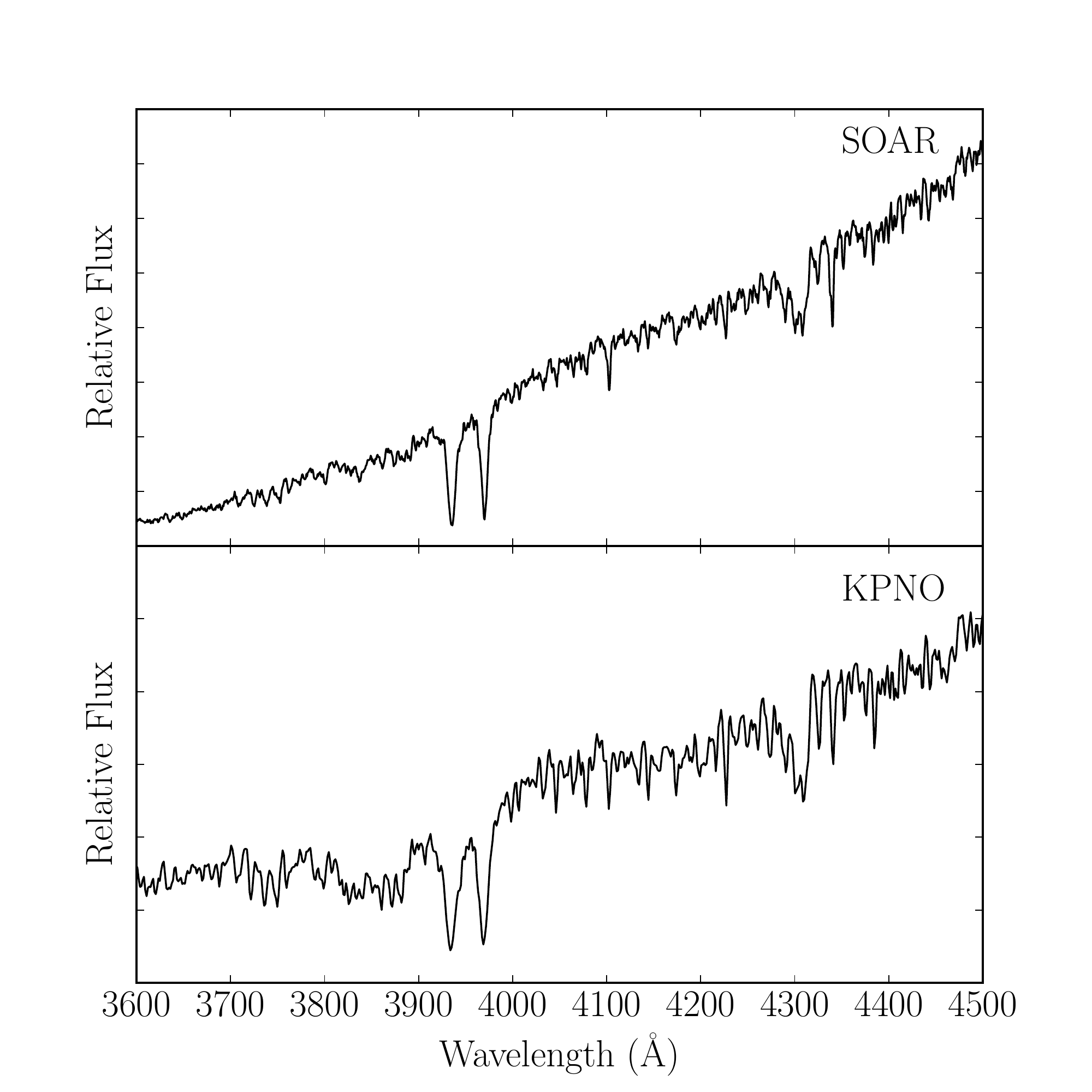}
\caption{Top Panel: a sample spectrum taken from the SOAR observations. The spectrum is 
 that of a giant from M30, a cluster with ${\rm [Fe/H]} = -2.3$. (\citealt{Harris}; 2010 edition)
Bottom Panel: a sample spectrum taken from the KPNO observations. The spectrum is 
 that of a giant from M10, which has an [Fe/H] value of $-1.56$. (\citealt{Harris}; 2010 edition)
 The CN and CH bands can be seen around 3800 \AA\ and 4300 \AA, respectively.}
\label{sample}
\end{figure}

Observations with the SOAR 4.1 m telescope were made under NOAO Program 
2011A-0352 during the three-day period of 2011 July 27-29. The Goodman High 
Throughput Spectrograph was utilized. The spectrograph was configured with
a 600 l mm$^{-1}$ diffraction grating, such that the resulting spectra have a 
resolution of 2.3 \AA\ FWHM and cover a wavelength range 
from the atmospheric violet cutoff to 6200 \AA. A slit width of 3'' was used. For each 
star, two to three exposures were taken to provide 1800 s of total exposure time. Poor 
weather conditions limited the observing time to 2.5 nights instead of the 
intended 3. The square root of the photons detected per pixel just red of the $\lambda$4300 CH band in 
each spectrum was taken to give an approximate S/N characterizing the spectra. Typical spectra had an S/N of $\sim50$ per pixel just blueward of the CH feature. A sample spectrum is shown as the bottom panel in Figure \ref{sample}.

\subsection{Reduction}

Spectra for each star from the KPNO observations were reduced following 
standard reduction procedures using the Image Reduction and Analysis Facility (IRAF).\footnote{IRAF is distributed by the 
National Optical Astronomy Observatory, which is operated by the Association 
of Universities for Research in Astronomy (AURA) under cooperative agreement 
with the National Science Foundation.} The CCD frames were bias and flat-field
corrected prior to the extraction of one-dimensional spectra. Initial 
wavelength calibrations were obtained from the comparison arc exposures
accompanying each stellar observation. There is a considerable range in
radial velocities among the clusters observed at KPNO. A radial velocity for 
each star was consequently calculated by cross-correlating the stellar spectrum
against a synthetic spectrum of a typical red giant to check for membership. 
Shifts obtained from the cross-correlation were used to place each spectrum in 
a stellar rest frame. Spectra were flux calibrated.

For the SOAR observations, a different method was used to reduce the spectra 
due to a large reflection in the flat-field images obtained from the telescope.
To eliminate the reflection, a series of flats were taken at an additional 
grating angle that moved the reflection to a different location on the detector. 
The region in the original flat with the reflection was then replaced with a 
section from the new flat after normalization.

Another issue encountered with the SOAR observations was that the arc 
lamp exposures for each the program were underexposed. Here a well-exposed 
arc taken at the start of the night was used to obtain the general shape of the pixel-wavelength solution, which
was then applied to the program stars. A cross-correlation between the 
resulting program spectrum and a synthetic template was then used to shift the 
observed spectra into the rest frame. A consequence of this approach was
the loss of any radial-velocity information. However, all SOAR program stars are believed to be members based on their positions on the color magnitude diagrams of the clusters.

\section{Analysis}

Carbon and nitrogen abundances were derived for each star by modeling two
spectroscopic indices measured from the spectra, one being a CH index
covering a strong CH absorption band located at 4300 \AA\ (the so-called $G$ 
band), with a second index chosen to be sensitive to a CN absorption feature 
with a bandhead at 3883 \AA. The CN index that was measured is denoted 
$S(3839)$ and was introduced by Norris et al. (1981) specifically for comparing
the intensity in the violet CN band at 3883 \AA\ with a nearby redward
comparison region of $\lambda\lambda$3883-3916 \AA\ that avoids CN absorption.
The $I{\rm(CH)}$ absorption index used employs two pseudo-continuum bandpasses,
a blueward region of 4240-4280 \AA, and a redward bandpass from 4390 to 4460 \AA.
The CH feature bandpass itself covers the wavelength range of 
$\lambda\lambda$4285-4315 \AA\ \citep{ch}. In the case of the $I{\rm(CH)}$ absorption index, each bandpass (continuum and feature) were normalized by their width by dividing the sum of the bandpass by its width in angstroms. Equations 1 and 2 show how the $S(3839)$ and $I{\rm(CH)}$ indices were calculated with $F_{X-Y}$ being the total integrated flux between wavelengths $X$ and $Y$.

\begin{equation}
S(3839) = - 2.5\log\frac{F_{3846-3883}}{F_{3883-3916}}
\end{equation}
\begin{equation}
I{\rm(CH)} = -2.5\log\frac{F_{4285-4315}/30}{0.5(C_1 + C_2)}
\end{equation}
\begin{equation}
C_1 = F_{4240-4280}/40
\end{equation}
\begin{equation}
C_2 = F_{4390-4460}/70
\end{equation}

The model atmospheres generated for each star were computed using the MARCS 
program \citep{MARCS} and the Synthetic Spectrum Generator \citep[SSG;][]{SSG,SSG2}.
Effective temperatures and surface gravities required for the stellar models 
were derived using the $V-K$ color of a star, the absolute magnitude, and an 
[Fe/H] value based on cluster membership. The apparent $V$ magnitudes for the 
program stars came from the sources referenced in Table 1, while the $K_s$ magnitudes
were taken from the 2MASS catalog \citep{2MASS}. The Harris catalog
(\citealt{Harris}; 2010 edition) was the source for the [Fe/H] adopted for each cluster. 

Determination of stellar temperatures and surface gravities was done according 
to the method described in \citet{Alonso} and \citet{Alonzo}. The equation in 
\citet{Alonzo} that gives the stellar effective temperature as a function of 
the $V-K$ color uses the Carlos Sanchez Telescope (TCS) photometric system. Thus 
$V-K$ colors from 2MASS were first converted to the TCS system using the method
described in \citet{Johnson}. The $V-K$ color on the TCS system was then 
corrected for reddening by subtracting the $E(B-V)$ values from \citet[][2010 edition]{Harris}
multiplied by 2.74, which is the adopted reddening ratio from \citet{V-K}.

\begin{figure}[htbp]
\centering
\includegraphics[trim = 0.4cm 0cm 0cm 0cm, scale=0.42, clip=True]{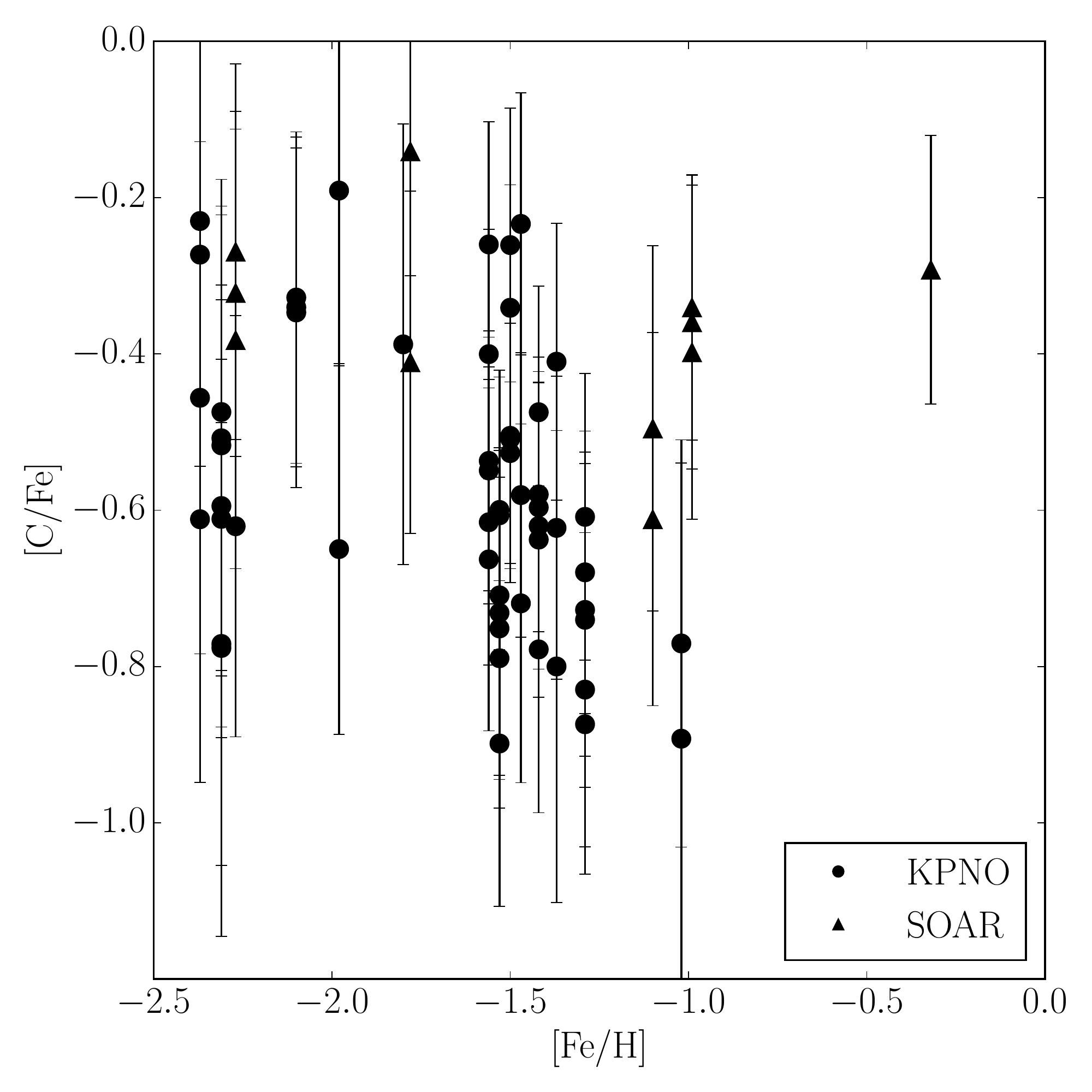}
\caption{Carbon abundance [C/Fe] vs. [Fe/H] for each star in the KPNO/SOAR
 sample. The circles indicate stars observed with the KPNO 4 m telescope and 
 triangles are those observed with the SOAR telescope. The error bars represent uncertainties from multiple observations of the same star. There is no clear 
 correlation between [C/Fe] and [Fe/H].}
\label{c-abundance}
\end{figure}

Synthetic spectra were calculated for each effective temperature and 
$\log g$ with the appropriate [Fe/H] then smoothed to match the observed 
spectrum for the individual stars. Carbon and nitrogen were adjusted 
simultaneously to match the CN and CH indices. Due to the role of the CO molecule in setting molecular equilibrium abundances in red giant photospheres, the [O/Fe] abundance used in generating the synthetic spectra for each star is an important constraint. Whenever possible, we used the literature values for any star in our study with measured [O/Fe] abundances \citep{sneden2004,cohen2005,Carretta,Carrett,johnson2012,boberg2015,boberg2016,rojasarriagada2016,villanova2016,marino2018}. If a star did not have a direct measurement, but belonged to a cluster with other [O/Fe] measurements, we assigned an [O/Fe] equal to the cluster average for that star. In the cases where no measurements were available for a cluster, an [O/Fe] abundance of 0.3 dex was assumed. This value is based on Figure 6 of \citet{Carretta} that shows [O/Fe] measurements for 19 GCs all approximately centered on 0.3 dex. 
$\xi = 2.0$ km s$^{-1}$ and $^{12}{\rm C}/^{13}{\rm C} = 4.0$ were also assumed 
for the synthetic spectra, which are reasonable estimates for GC RGB stars \citep[e.g.,][]{suntzeff1991,pavlenko2003}. The resulting carbon abundances are plotted in 
Fig. \ref{c-abundance} as a function of the metallicity of each cluster. The derived [C/Fe] 
and [N/Fe] values are presented in Table \ref{abundance}. 

To evaluate our uncertainties of the resulting abundances based on the sensitivity to [O/Fe], [Fe/H], the
$^{12}{\rm C}/^{13}{\rm C}$ ratio, and errors in temperatures, gravities, and band strengths, we repeated the C and N abundance determinations using different 
values for these input parameters as shown in Table \ref{changes}. We chose to vary [O/Fe] by $\pm$ 0.3 dex as this range encompasses the [O/Fe] of most stars observed by \citet{Carretta,Carrett}. Maximum uncertainties for effective temperatures based on the method in \citet{Alonso} and \citet{Alonzo} are 150 K, which results in uncertainties in surface gravity of 0.2, so we varied these parameters by these values. For the band strength uncertainties, we subtracted the uncertainty in $I(\rm{CH})$ and added the uncertainty in $S(3839)$ to maximize the effect of these uncertainties since the bands are negatively correlated. A final uncertainty for each star is then calculated by adding the changes from each of these assumptions in quadrature and provided in Table \ref{abundance}. The average change on [C/Fe] and [N/Fe] from each of these parameters for the sample is also 
given in Table \ref{changes}, which shows that the greatest sensitivity is to $T_{\rm{eff}}$ and [O/Fe] and demonstrates the importance of our careful selection of [O/Fe] values. 

A further check on our methods for assigning parameters such as O abundances comes from comparisons with the study of \citet{Martell}. We note that our study has nine stars in common with \citet{Martell} who determined carbon abundances for RGB stars in GCs through a similar method to ours (matching CN and CH bands with synthetic spectra). We find that for eight of these nine stars our carbon abundances agree within the uncertainties. The one star that does not agree was likely the result of a poor observation by \citet{Martell} as it was left out of the main analysis of that paper. This comparison and consistency between the two studies provides additional support that our methods for determining parameters for each star are appropriate.

\begin{deluxetable*}{ccccccccccc}
\tabletypesize{\footnotesize}
\tablewidth{0pt}
\tablecaption{Abundances for program stars\label{abundance}}
\tablehead{\colhead{Globular Cluster} & \colhead{[Fe/H]} & \colhead{$V$} & 
\colhead{$M_\mathrm{V}$} & \colhead{$T_{\rm eff}$ (K)} & \colhead{log($g$)} & \colhead{$S(3839)$} & \colhead{$I{\rm(CH)}$} & \colhead{[C/Fe]\tablenotemark{a}} & \colhead{[N/Fe]\tablenotemark{a}} &
\colhead{Telescope}}
\startdata
NGC 6362 (1)& & & & & & & & & \\
\hline
4&-0.99&13.32&-1.36&4328&1.36&0.311&0.389&-0.34 $\pm$ 0.17&0.79 $\pm$ 0.12&SOAR\\
6&-0.99&13.23&-1.45&4216&1.24&0.483&0.402&-0.36 $\pm$ 0.19&1.10 $\pm$ 0.24&SOAR\\
25&-0.99&13.32&-1.36&4108&1.19&0.228&0.408&-0.40 $\pm$ 0.21&0.78 $\pm$ 0.24&SOAR\\
NGC 6723 (2)& & & & & & \\
\hline
1-5&-1.1&13.30&-1.54&3942&0.97&0.432&0.396&-0.61 $\pm$ 0.24&1.76 $\pm$ 0.3&SOAR\\
2-4-62&-1.1&13.38&-1.46&4004&1.06&0.414&0.400&-0.50 $\pm$ 0.23&1.42 $\pm$ 0.25&SOAR\\
NGC 7099 (M30) (3), (4)& & & & & & \\
\hline
PE-19&-2.27&13.04&-1.6&4537&1.39&0.037&0.255&-0.27 $\pm$ 0.24&1.41 $\pm$ 0.10&SOAR\\
38&-2.27&13.24&-1.4&4642&1.53&-0.027&0.189&-0.38 $\pm$ 0.29&1.56 $\pm$ 0.11&SOAR\\
157&-2.27&13.00&-1.64&4419&1.30&-0.014&0.278&-0.32 $\pm$ 0.21&1.15 $\pm$ 0.10&SOAR\\
\enddata
\tablenotemark{Based on solar abundances from \citet{asplund2009}}
\tablecomments{This table is available in its entirety in machine-readable form.}
\tablerefs{(1)\citet{ALC}, (2)\citet{Menzies}, 
(3)\citet{Alcaino}, (4)\citet{Dickens} }
\end{deluxetable*}

\begin{deluxetable}{ccc}
\tabletypesize{\footnotesize}
\tablewidth{0pt}
\tablecaption{Sensitivity of the derived [C/Fe] abundance to input parameters for
SSG synthetic spectra \label{changes}}
\tablehead{\colhead{Parameter Change} & \colhead{Average $\Delta$[C/Fe]} & \colhead{$\Delta$[N/Fe]}}
\startdata
$T_{\rm eff} \pm 150$ K & 0.16 & 0.12\\
log(g) $\pm$ 0.2 & 0.03 & 0.04\\
$[$O/Fe$]$ $\pm$ 0.3 dex & 0.10 & 0.05\\
$[$Fe/H$]$ $\pm$ 0.1 dex & 0.05 & 0.03\\
$S(3839)$ and $I{\rm(CH)}$ $\pm$ 0.01 & 0.05 & 0.08\\
${^{12}{\rm C}}/{}{^{13}{\rm C}}$ $\sim + 10$ & 0.02 & 0.02\\
\enddata
\end{deluxetable}

\section{Results}

A summary of the carbon abundances derived in this program is shown in 
Fig.~\ref{c-abundance}, which is a plot of [C/Fe] versus [Fe/H] for each star in the sample 
(circles for stars observed at KPNO; triangles for stars observed at SOAR). 
All but one star has ${\rm [C/Fe]} < -0.2$, with some being carbon depleted
by as much as ${\rm [C/Fe]} \sim -1.0$. Such non-solar [C/Fe] ratios are signs that carbon has likely been depleted at the surfaces of the cluster giants (halo subdwarfs, for example, more typically have near-solar [C/Fe]; see for example, \citet{Laird}, \citet{Carbon}, \citet{GSCB}). 
If GC stars were formed with initial carbon abundances 
commensurate with those of subdwarfs then the implication of the abundances in
Fig.~\ref{c-abundance} is that a process has been at work within the cluster giants that
has brought about a reduction in the surface carbon abundance. Within the 
context of the interior structure of the GC giants, some physical
mechanism has caused carbon-depleted material from the CNO-bi-cycle 
hydrogen-burning shell to be transported across the radiative region that
surrounds the shell and thence carried into the convective envelope, whereupon 
it is rapidly convected to the surface of the red giant. We colloquially use 
the term extra mixing to refer to the process that produces these carbon 
depletions. This refers to a non-convective mixing or mass transport of 
material across the interior radiative zone of the red giant, i.e., it is a
mixing process that acts in addition to the outer convective envelope. 
Although there may seem to be no pattern to the distribution of points in 
Fig.~\ref{c-abundance}, there is an underlying dependence of deep mixing on metallicity implicit
in these data.

Stellar interior theory indicates that mixing between the base of the 
convective envelope of a cluster giant and the outer parts of the
hydrogen-burning shell can only take place after this shell has burned through 
a molecular weight discontinuity left within the star at a location 
corresponding to the deepest inward extent of the base of the convective 
envelope during evolution up the RGB \citep{Iben,Ibe}. 
The occurrence of this burn through event produces a local maximum in the 
RGB luminosity function. Taking this conventional theory as being correct, 
for each of the clusters in our sample the absolute magnitude of the LMLF can 
be predicted. Through further application of stellar evolution models the time 
since each star in our sample has evolved through the LMLF can be calculated. 
Assuming a primordial [C/Fe] for each cluster, a comparison with the observed 
[C/Fe] values then allows the calculation of a carbon depletion rate on the 
RGB to be derived for each star. A depletion rate is defined 
here as the ratio of the change incurred in the [C/Fe] abundance  
to the time since a star evolved through the LMLF. Such calculated depletion 
rates can be correlated with metallicity - the prime goal of our project.

\begin{figure}[htbp]
\centering
\includegraphics[trim = 0.4cm 0cm 0cm 0cm, scale=0.42, clip=True]{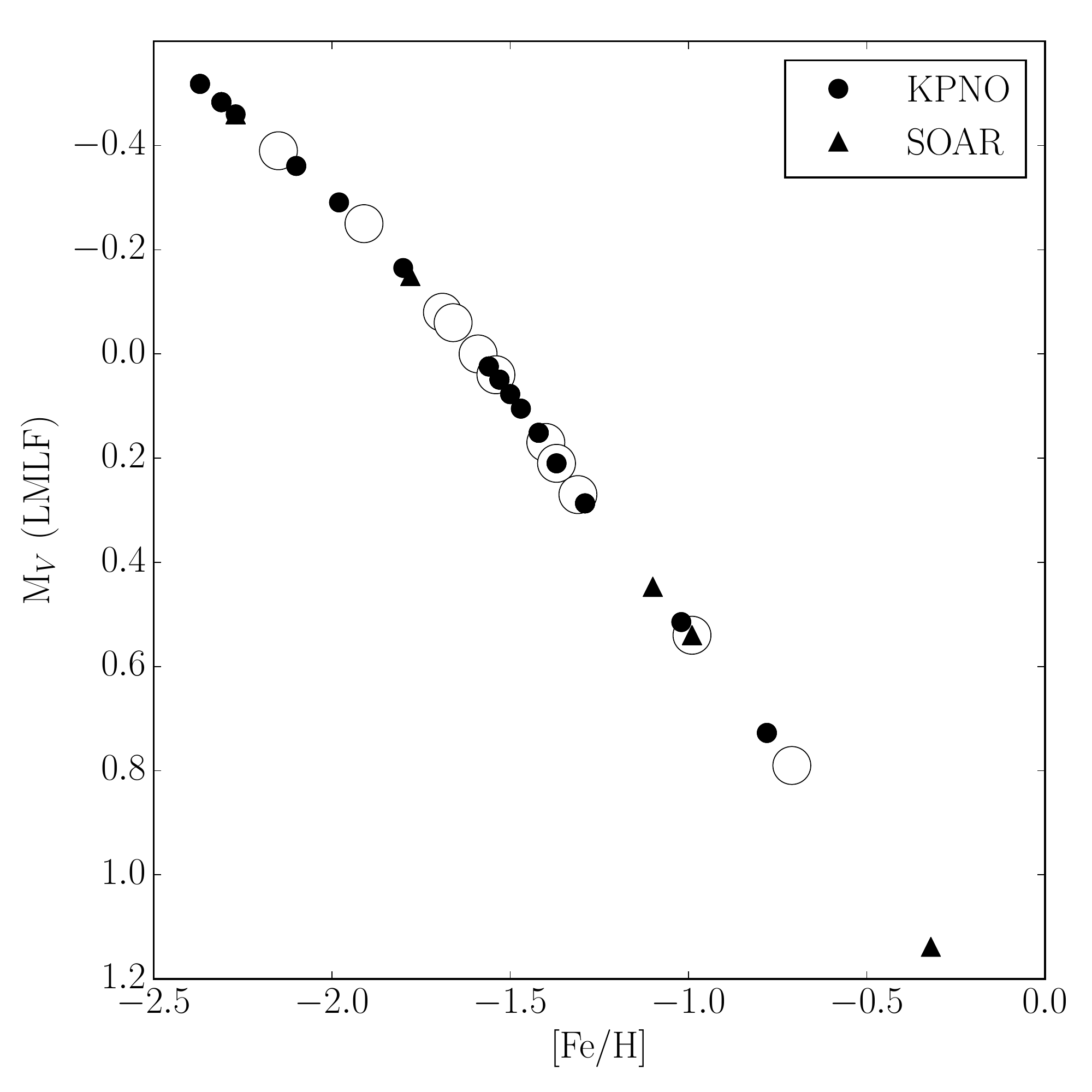}
\caption{Absolute magnitude of the local maximum in the luminosity 
 function of a globular cluster as a function of metallicity. The data points
 are taken from \citet{bump}. The solid circles represent clusters 
 observed as part of the KPNO sample, and the triangles represent clusters 
 observed with the SOAR telescope. Clusters from \citet{bump} not  
 in our sample are shown as open circles. The solid lines are linear 
 interpolations between the data points.}
\label{mvbump}
\end{figure}

The absolute magnitude of the LMLF for each cluster was estimated using a 
correlation between this magnitude and cluster metallicity that was derived
from the cluster data set of \citet{bump}. The correlation is shown
in Fig. \ref{mvbump}. A linear interpolation was used with the data from \citet{bump}
to calculate a value of $M_\mathrm{V}$(LMLF) for each cluster in our 
sample. The LMLF magnitudes from \citet{bump}, as well as the 
values derived for the clusters in our sample, are both shown in Fig. \ref{mvbump}.

Following the method of \citet{Martell}, we use Yale Y$^2$ isochrones \citep{YY} to determine a characteristic 
rate for the luminosity evolution, $\Delta M_\mathrm{V} / \Delta t$, of stars with 
metallicities corresponding to each GC in the observational
program. A stellar model for a given [Fe/H] was chosen within 0.05 mag of the 
LMLF from an isochrone calculated with an age of 12 Gyr. This model sets the 
mass of a 12-Gyr-old star that is evolving through the LMLF for a cluster of
the given [Fe/H]. Additional isochrones of greater age were then reviewed 
until one was found where a model of the appropriate mass had evolved to an 
absolute magnitude of $M_\mathrm{V} = -1.5 \pm 0.05$. The age difference, $\Delta t$
between the two isochrones provided a value for the ratio 
$\Delta M_\mathrm{V} / \Delta t$, representing the mean rate at which
a star evolves up the RGB past the LMLF in each cluster.

Once a change in magnitude as a function of time had been calculated for each 
star, a change in carbon abundance as a function of magnitude was determined 
from (i) the difference in [C/Fe] from an assumed initial solar [C/Fe] and (ii) the difference in absolute magnitude of the star from the LMLF of the cluster. With $\Delta {\rm [C/Fe]} / \Delta M_\mathrm{V}$ thereby established for each star in the sample, a value for the carbon depletion rate, $\Delta {\rm [C/Fe]} / \Delta t$, was derived with units of dex per gigayear.

\begin{figure}[htbp]
\begin{center}
\includegraphics[trim = 0.4cm 0cm 0cm 0cm, scale=0.42, clip=True]{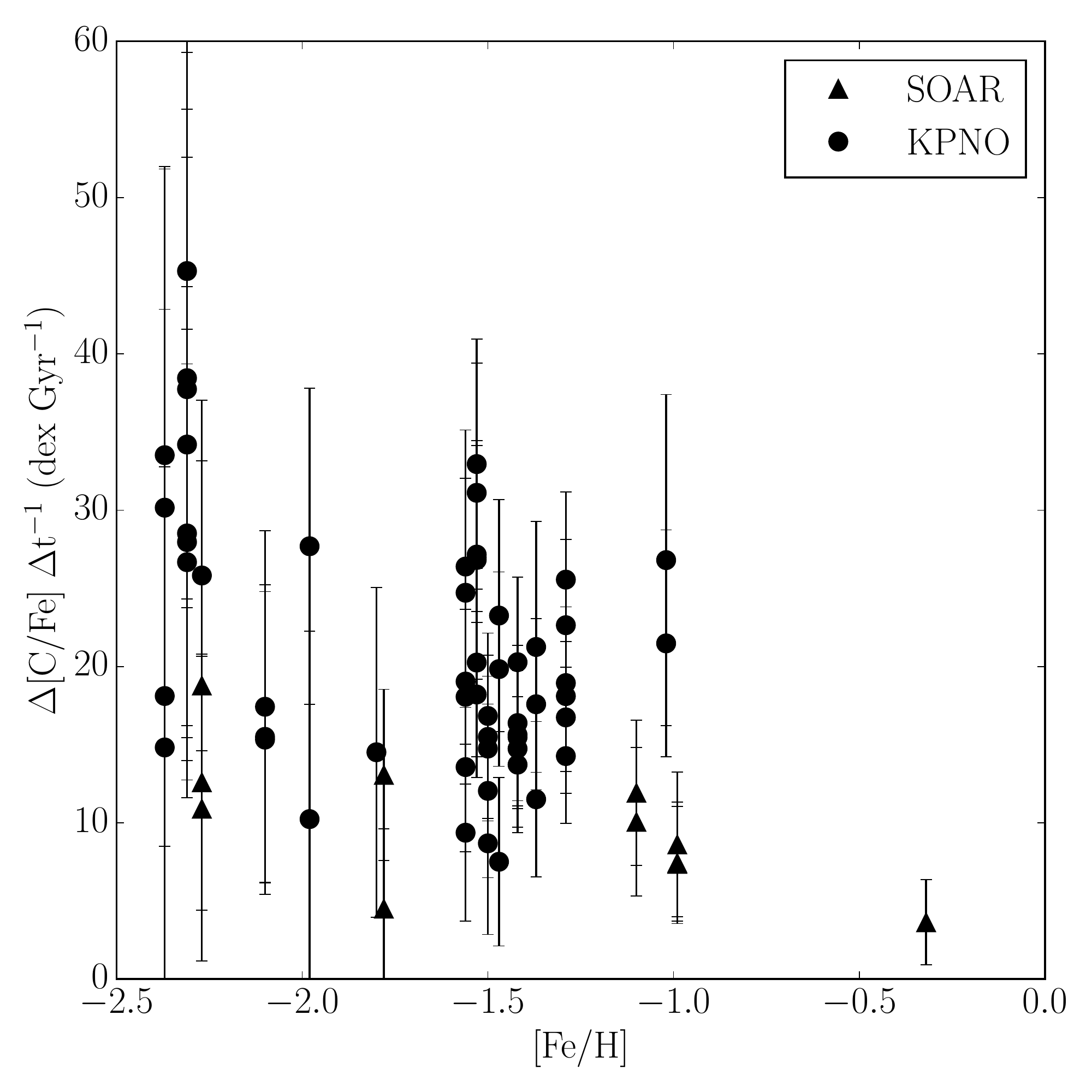}
\caption{Inferred carbon depletion rate as a function of [Fe/H] for each 
 member of our sample. Members from the SOAR observations are shown as triangles 
 and the KPNO observations are shown as circles. To obtain this diagram the 
 assumption has been made that all stars formed with the same initial solar 
 [C/Fe], which is later modified by deep mixing. A general decrease in the carbon 
 depletion rate is seen with increasing [Fe/H] metallicity, although there is 
 considerable scatter (especially around ${\rm [Fe/H]} \sim -1.5$).} 
\label{carbon-rate-plot}
\end{center}
\end{figure}

A plot of the carbon depletion rate for each star as a function of metallicity
is shown in Fig.~\ref{carbon-rate-plot} with error bars showing the propagated uncertainty that was calculated for our [C/Fe] measurements from Section 3. A correlation with [Fe/H] is evident, and in this sense 
our result is analogous to that of \citet{Martell}, whose methodology
we have followed so far. However, there is a 
large spread in the calculated depletion rate at each metallicity, especially 
around ${\rm [Fe/H]} \sim -1.5$ where the sample is most numerous. 

One complication is that in some clusters it is known that there is a 
relation between CN and CH band strengths among red giants of similar
abundance magnitudes \citep[][and references therein]{gratton}. This translates into a scatter in [C/Fe] at a 
given $M_\mathrm{V}$ in such a way that the [C/Fe] and [N/Fe] are anticorrelated, with
CN-strong giants having [C/Fe] abundances lower than those of CN-weak giants
of comparable luminosity. Thus one might expect to derive different values of
$\Delta {\rm [C/Fe]} / \Delta M_\mathrm{V}$ for CN-strong and CN-weak giants. To make 
allowance for this scatter, [N/Fe], having been derived by simultaneously 
fitting the $\lambda$3883 CN index, was noted for each star in order to assign
it to a nitrogen-rich (CN-strong) or nitrogen-normal (CN-weak) category. If a 
star is nitrogen-rich, it is taken here to indicate that the star formed from 
gas that incorporated CN(O)-cycled material through some poorly understood process very early in the cluster history. While the 
initial CN(O)-cycled material was nitrogen-enhanced, it also had been depleted in 
carbon; therefore, stars that were formed as nitrogen-rich would generally have
had a lower initial carbon abundance than nitrogen-normal stars within the same
cluster. An analysis of a histogram of [N/Fe] for the sample showed two 
peaks clearly separated at [N/Fe] $\sim$ 1.1. Therefore, stars with [N/Fe] greater than 1.1 in this sample were considered nitrogen-rich, and those with
[N/Fe] less than 1.1 were considered nitrogen-normal.

\begin{figure}[htbp]
\begin{center}
\includegraphics[trim = 0.4cm 0cm 0cm 0cm, scale=0.42, clip=True]{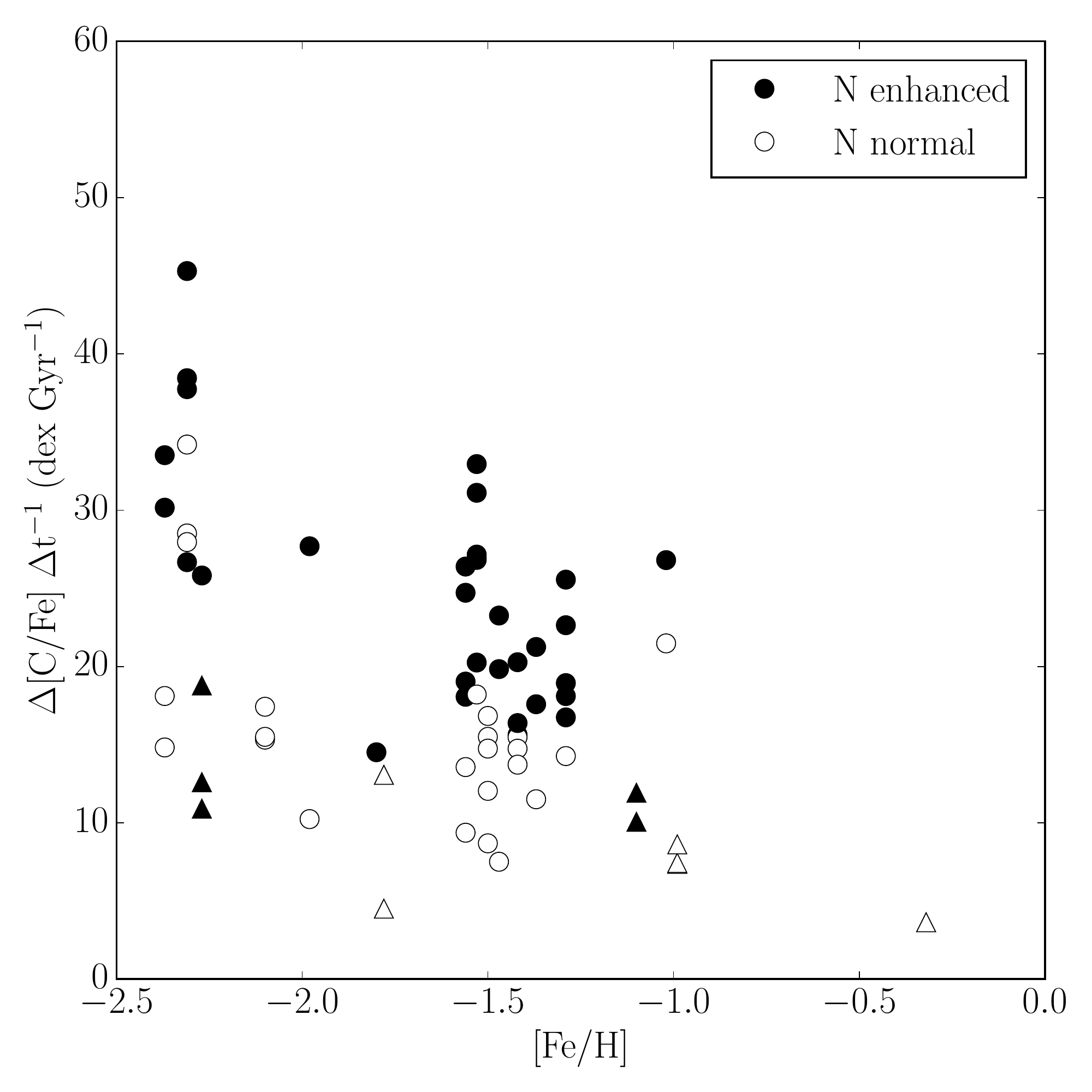}
\caption{Carbon depletion rate vs. metallicity. Analogous to Fig.~\ref{carbon-rate-plot}, except that nitrogen-normal stars are depicted as open circles for the KPNO data and open triangles for the SOAR data. Likewise, filled circles and triangles depict nitrogen-enhanced stars observed at these two telescopes. \textbf{The error bars are removed for clarity, but are the same as in Fig. \ref{carbon-rate-plot}.}}
\label{carbon-rate-split}
\end{center}
\end{figure}

\begin{figure}[htbp]
\begin{center}
\includegraphics[trim = 0.4cm 0cm 0cm 0cm, scale=0.42, clip=True]{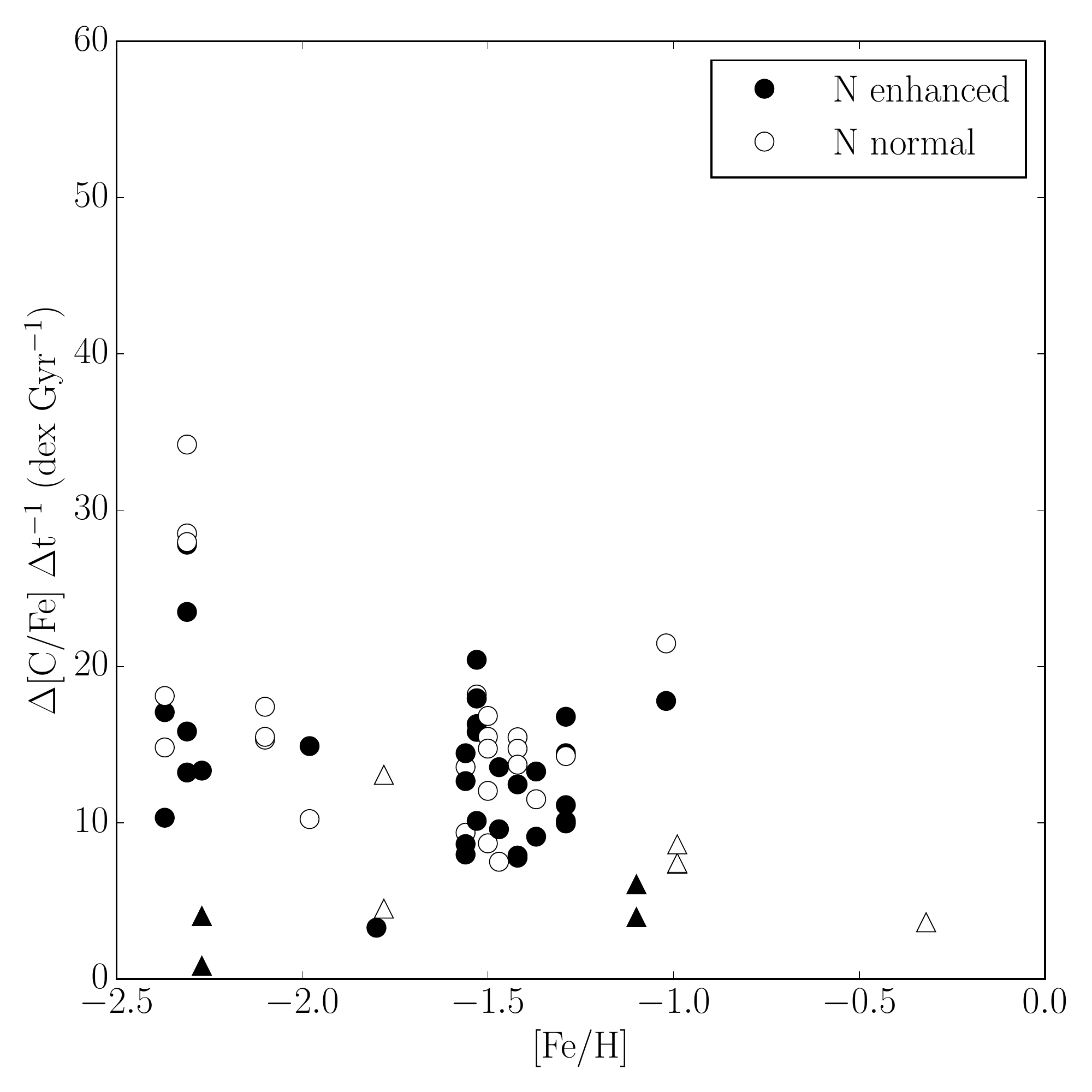}
\caption{Carbon depletion rate as a function of metallicity after allowing for
 initial differences in [C/Fe] between nitrogen-rich and nitrogen-normal 
 giants. The carbon depletion rates for the nitrogen-rich giants have been 
 lowered to account for their lower initial carbon abundance. The spread in 
 carbon depletion rates within each cluster is reduced, especially around  
 ${\rm [Fe/H]} \sim -1.5$.}
\label{carbon-rate-adj}
\end{center}
\end{figure}

A plot of carbon depletion rate versus metallicity is shown in Fig.~\ref{carbon-rate-split} with the stars indicated as nitrogen-rich or nitrogen-normal. In general, it appears as though nitrogen-rich giants have higher carbon depletion rates than nitrogen-normal ones. However, the carbon depletion rates are likely overestimated for the nitrogen-rich stars because their initial carbon abundances are likely lower than assumed in the calculation of $\Delta$[C/Fe]. If the initial [C/Fe] is lowered to $-0.3$ for the nitrogen-rich stars, the resulting
$\Delta$[C/Fe] is reduced as well as the calculated carbon depletion rate. 
Lowering the carbon depletion rates in the nitrogen-rich stars according to 
this recipe results in values of $\Delta {\rm [C/Fe]}/ \Delta t$ that are 
plotted versus [Fe/H] in Fig.~\ref{carbon-rate-adj}. The scatter in depletion rate at a given [Fe/H] is reduced as compared with the spread seen in 
Figure~\ref{carbon-rate-plot}, particularly 
around stars in clusters with a metallicity of ${\rm [Fe/H]} \sim -1.5$.

Comparison between Figs. \ref{carbon-rate-plot} and \ref{carbon-rate-adj} illustrates the degree to which the inferred
carbon depletion rates are dependent upon assuming an initial [C/Fe] for the
red giants in our cluster sample. To precisely determine
$\Delta {\rm [C/Fe]}/ \Delta t$ for a given RGB star, one would need to 
know the initial [C/Fe] for that star before it began deep mixing. Determining 
[C/Fe] and [N/Fe] for stars on the main sequence of a cluster, or just before 
the point of mixing (the subgiant or early giant phase), could provide an 
estimate for the initial [C/Fe] for nitrogen-rich RGB stars. However, recent 
studies have shown that in some clusters the [C/Fe] and [N/Fe] abundances of
main sequence and early turn-off stars spread over a wide range 
($\sim 0.5$ dex) rather than forming a bimodal distribution that could
conveniently be identified with nitrogen-rich and nitrogen-normal abundances 
(e.g., \citet{Briley,Cohen}). Because of the range in initial [C/Fe] and 
[N/Fe] abundances before mixing has begun, a value for the initial [C/Fe] of 
any particular star currently in a deep mixing phase on the upper RGB may be
difficult to determine. Such uncertainty will inevitably limit the precision 
with which $\Delta {\rm [C/Fe]}/ \Delta t$ can be determined through the
type of technique that we have employed in this observational program. The more
robust, but more telescope time intensive, approach to deriving
$\Delta {\rm [C/Fe]}/ \Delta t$ for a particular cluster would be to measure
$\Delta {\rm [C/Fe]}/ \Delta M_\mathrm{V}$ for CN-strong and CN-weak giants separately
spanning a large luminosity range on the RGB, similar to what has been done by \citet{suntzeff1981}, \citet{carbon1982}, \citet{trefzger1983},
\citet{langer1986}, \citet{bellman2001}, \citet{shetrone2010}, \citet{kirby2015}, and \citet{Gerber} for various clusters.

\section{Conclusions}
We have presented our findings for carbon depletion rates among GC red giants for a range of metallicities following a method similar to \citet{Martell}. While we were unable to reduce the star-to-star scatter observed by that work, we were able to determine that the scatter is caused by more than variability in mixing rates. We found that this method of determining mixing rates is heavily dependent on the assumed initial carbon abundance for each star. \citet{Laird}, \citet{Carbon}, and \citet{GSCB} have shown that a solar [C/Fe] value is a reasonable assumption for the initial abundance based on values measured in halo subdwarfs, but the multiple populations now known to exist in most Galactic GCs \citep[see, e.g.,][]{piotto2015} add complications since the initial carbon abundance will be dependent on the population a RGB star belongs to.

Taking this into consideration, we find the typical rate of carbon depletion to range from $\sim$20 to 50 dex Gyr$^{-1}$, with higher rates being found among metal-poorer clusters. Extra mixing has been assumed here to commence when a star evolves through the magnitude corresponding to the LMLF. In the clusters of our sample, the typical time taken to evolve from the $M_\mathrm{V}$(LMLF) to the tip of the RGB is $\sim$30-50 Myr (depending on the cluster, with lower metallicity clusters taking less time). If mixing were to continue at the same average rate throughout the entire upper RGB, the total amounts of carbon depletion incurred by GC red giants just prior to the core-helium flash would be $\sim$1.0-1.5 dex dependent upon metallicity. Thus some quite substantial carbon depletion would be anticipated among GC stars as they evolve onto the horizontal branch. We also note that the complications from multiple populations mean that more observationally intensive methods that measure large samples of red giants in each cluster will be needed to separate changes caused by deep mixing from those caused by multiple populations.

\FloatBarrier

\section{Acknowledgements}

We would like to thank Roger A. Bell for making the SSG program available to 
us. M.~M. Briley and G.~H. Smith gratefully acknowledge support by the 
National Science Foundation through the grants AST-0908924 and 
AST-0908757, respectively.

This publication makes use of data products from the Two Micron All Sky Survey, which is a joint project of the University of Massachusetts and the Infrared Processing and Analysis Center/California Institute of Technology, funded by the National Aeronautics and Space Administration and the National Science Foundation.

\FloatBarrier

\end{document}